\documentclass[11pt]{article}
\usepackage[margin=1in]{geometry}
\usepackage{amsmath,amssymb}
\usepackage{graphicx}
\usepackage{booktabs}
\usepackage{xcolor}
\usepackage[hidelinks]{hyperref}
\usepackage{caption}
\title{\textbf{Dynamical and Optimization Trade-offs of\\ Levi--Civita Coordinates for Learned\\ Close-Encounter Dynamics}}
\author{Abhishek Shankar\\[2pt]
openproblem.ai\\[2pt]
\texttt{abhishek.shankar@openproblem.ai}\\[2pt]
ORCID: \href{https://orcid.org/0009-0002-6552-3622}{0009-0002-6552-3622}}
\date{}

\begin{document}
\maketitle

\begin{abstract}
\noindent
Classical regularization removes the binary-collision singularity from the
Kepler problem, but its value as a \emph{representation} for learned Hamiltonian
dynamics has not been systematically isolated. We compare Cartesian and planar
Levi--Civita formulations of a perturbed Kepler system with a smooth quadrupole
potential. With the perturbation supplied analytically, a Levi--Civita
Hamiltonian splitting holds the maximum relative energy error near
$2.1\times10^{-5}$ through eccentricity $e=0.99$ (a discretization baseline
converging as $\mathcal{O}(\Delta s^2)$, not a ceiling), while the Cartesian
splitting becomes unstable. This advantage survives a matched-horizon,
matched-budget control --- at equal physical horizon $T_0$ and equal
force-evaluation count the regularized baseline is $3\times10^{-5}$, approximately
$4.7$--$8.3$ orders of magnitude below the Cartesian arm depending on eccentricity. In
held-out high-eccentricity tests with matched
physical-time sampling, the regularized models produce finite rollouts in $40/40$
runs versus $0/40$ for Cartesian --- though in a fixed-shell formulation that
supplies the regularized model with the exact initial orbit energy, an asymmetry
we disclose but do not yet control, and though ``survival'' at these settings
still carries $\mathcal{O}(1)$ energy error (survival, not accuracy). Four neural
residual objectives fail to approach the analytic result. We trace this with
exact-feature controls: the regularized residual is exactly a four-monomial
degree-6 polynomial that a direct least-squares solve fits to the baseline, so
the \emph{exact-feature} gap is one of optimization geometry (severe raw-basis
ill-conditioning that orthogonalization removes for L-BFGS in two iterations),
not nonrepresentability. We therefore report a \emph{trade-off} between
dynamical/numerical conditioning, which regularization improves, and
regression/optimization conditioning, which the raw transformed basis worsens.
The \emph{separate} small-MLP gap remains unexplained (a gauge-invariant variant
does not close it). This is a controlled falsification-plus-trade-off study, not a
solution to learned close-encounter dynamics.
\end{abstract}

\section{Introduction}
A \emph{physics-native} model of gravitational dynamics aims to combine a neural
parameterization of the interaction with structural guarantees --- energy and
momentum conservation, symplecticity --- so that predictions remain physical
over long horizons. The dominant design pattern is a Hamiltonian neural network
(HNN) or Hamiltonian graph network whose scalar $H_\theta(q,p)$ is differentiated
to yield forces, integrated by a symplectic scheme. Most neural
celestial-dynamics benchmarks we identified operate in physical Cartesian states
or orbital-element representations rather than analytically regularized
Levi--Civita/Kustaanheimo--Stiefel phase space; HNN research does also include
canonical generalized, latent, and learned coordinate systems.

Cartesian coordinates carry a structural liability that is specific to
gravitation and absent from the pendulum/spring/molecular benchmarks the field
grew up on: the two-body potential $-\mu/r$ and its force $-\mu\, q/r^3$ are
singular at $r\to0$. Near a close encounter the force spans orders of magnitude
over a short arc, which (i) makes the target ill-conditioned for a
fixed-capacity network at a fixed training budget, and (ii) forces any
fixed-step integrator to take prohibitively small steps or diverge.

Classical celestial mechanics solved this a century ago by \emph{regularization}:
a coordinate and time transformation that removes the collision singularity
analytically. In the planar Levi--Civita map $q=z^2$ (with $z=u+iw$) together
with the Sundman time transform $dt=r\,ds$, the Kepler problem becomes a
\emph{harmonic oscillator} in the fictitious time $s$~\cite{levicivita1920,stiefel1971}.
This is exact and classical. We are not aware of prior work that combines an
analytic Levi--Civita Kepler core, a learned regularized residual potential, and
symplectic splitting specifically to study neural close-encounter dynamics,
though adjacent ideas exist: learned invertible coordinate transforms that
enforce dynamics constraints in a transformed frame~\cite{unifying2022}, and deep
Koopman models of perturbed two-body and restricted three-body
motion~\cite{koopman2024}. Our positioning claim is correspondingly narrow: we
identify this specific combination --- an analytic regularizing map with a learned
smooth, nonsingular residual (over the sampled compact domain) and a close-encounter symplectic rollout --- as the contribution,
not the individual components.

This paper asks the narrow, falsifiable question: \emph{what numerical benefit
does Levi--Civita regularization provide to learned close-encounter models, and
what optimization cost does the transformed residual introduce?} The answer has
several parts: the analytic architecture removes the numerical close-encounter
instability (yes); the learned formulation improves bounded rollout survival but
not accuracy; the raw transformed polynomial basis introduces an optimization cost
that the controls attribute to raw-basis conditioning; and accurate neural
residual learning remains unresolved.

\paragraph{Contributions.}
\begin{enumerate}
\item We show that a gray-box separable Hamiltonian (analytic harmonic Kepler
core $+$ residual) with a symplectic integrator, in Levi--Civita coordinates,
holds relative energy error at ${\sim}2.1\times10^{-5}$ through $e=0.99$ while
the Cartesian equivalent becomes numerically unstable; we show by a step-size
sweep that this value is the \emph{$N{=}400$ discretization baseline} converging
as $\mathcal{O}(\Delta s^2)$, not an architectural ceiling. The advantage persists
when both representations are integrated to the same physical horizon $T_0$ with
the same $400$-step force-evaluation budget (Section~\ref{sec:arch}).
\item In a controlled multi-seed ablation with time-step adaptivity held fixed,
the regularized formulation produces a large high-eccentricity rollout-survival
advantage under matched sampling and compute: the Cartesian formulation diverges
in $40/40$ in-distribution high-$e$ close encounters, while the regularized
formulation diverges in $0/40$ (Section~\ref{sec:ablation}). We flag that its
fixed-shell construction additionally supplies the exact orbit energy, an
information asymmetry we treat as a confound rather than a clean coordinate
isolation.
\item We separate the \emph{exact-feature} optimization gap from the unresolved
\emph{neural} approximation gap (Section~\ref{sec:negative}). For the exact
polynomial basis, the residual is a four-monomial degree-6 polynomial that a
direct linear solve fits to the baseline; for the same underlying physical
perturbation represented in the two coordinate systems, iterative
optimizers reach the baseline in \emph{physical} coordinates but the raw
\emph{Levi--Civita} basis is severely ill-conditioned and QR orthogonalization
restores baseline fitting for L-BFGS in two iterations. Small tanh MLPs, however,
remain at $\mathcal{O}(1)$ rollout error even after gauge symmetrization, and
their failure remains open. The Levi--Civita transform therefore trades
\emph{improved dynamical conditioning} for \emph{worsened regression conditioning
of this raw polynomial parameterization} --- the paper's central finding --- while
the neural gap is left explicitly unexplained.
\end{enumerate}

\paragraph{What we do \emph{not} claim.}
We do not claim that regularized coordinates are a novel idea, that this is a
working world model, or that the residual-learning gap is a fundamental
bottleneck. The system is
two-body-dominated with a fixed regularization chart and a smooth analytic
perturbation; the orbit energy $E$ entering the regularized Hamiltonian is
computed analytically at the initial state (Section~\ref{sec:setup}), so the
learned experiments are not fully self-contained and the survival ablation
carries an $E$-information confound (Section~\ref{sec:ablation}). We do not
benchmark against production integrators (REBOUND/WHFast, IAS15, TRACE, chain
regularization), which remain the correct tools for the forward problem, and we
report a single conserved quantity (energy); trajectory-level metrics are left
to future work.

\section{Setup}
\label{sec:setup}
We work with the planar perturbed Kepler problem in units $\mu=1$:
\begin{equation}
H(q,p)=\tfrac12\lVert p\rVert^2-\frac{\mu}{\lVert q\rVert}+V_{\mathrm{pert}}(q),
\end{equation}
with a \emph{smooth} tidal-quadrupole perturbation
$V_{\mathrm{pert}}(x,y)=A\,(x^2-\tfrac12 y^2)$, $A=0.02$. Its gradient
$\nabla V_{\mathrm{pert}}=(2Ax,-Ay)$ is smooth and bounded over the compact state
domain sampled in our experiments (it grows linearly at large $\lVert q\rVert$,
but the bound orbits never reach that regime); crucially it is \emph{non-singular}
at $r\to0$, so the residual difficulty is not a singularity artifact.

\paragraph{Regularized coordinates.}
With $z=u+iw$, the Levi--Civita map is $x=u^2-w^2$, $y=2uw$, $r=u^2+w^2$, and
the Sundman transform $dt=r\,ds$. Writing $P=4z'$ for the conjugate momentum in
$s$, the regularized (Poincar\'e) Hamiltonian on the energy shell $H=E$ is
\begin{equation}
\Gamma=\frac{\lVert P\rVert^2}{8}+\lvert E\rvert\,(u^2+w^2)-\mu+U_{\mathrm{res}},
\qquad U_{\mathrm{res}}=r\,V_{\mathrm{pert}},
\label{eq:gamma}
\end{equation}
which is \emph{separable}: the Kepler part is an exact harmonic oscillator
$\lvert E\rvert(u^2+w^2)$, and all perturbation content sits in $U_{\mathrm{res}}$.
The general fixed-shell Poincar\'e term is $-E\,r$; all trajectories considered
here are bound ($E<0$), so $\lvert E\rvert=-E$ and the oscillator stiffness
$\lvert E\rvert>0$ is written with the absolute value for clarity.
The map is completed by the canonical momentum transformation and its inverse,
which we state explicitly so the experiment is reproducible without the code. With
$P=P_u+iP_w$ the momentum conjugate to $z=u+iw$ and $p=(p_x,p_y)$ the physical
momentum, and $r=u^2+w^2$,
\begin{equation}
\begin{pmatrix}P_u\\P_w\end{pmatrix}
=2\begin{pmatrix}u & w\\ -w & u\end{pmatrix}\begin{pmatrix}p_x\\p_y\end{pmatrix},
\qquad
\begin{pmatrix}p_x\\p_y\end{pmatrix}
=\frac{1}{2r}\begin{pmatrix}u & -w\\ w & u\end{pmatrix}\begin{pmatrix}P_u\\P_w\end{pmatrix},
\end{equation}
equivalently $P=2\bar z\,p$ and $p=zP/(2r)$ in complex form; the Sundman
relation is $dt/ds=r$, integrated alongside the equations of motion, and initial
conditions are chosen on the shell $\Gamma(u_0,w_0,P_0)=0$ by evaluating $E$
analytically at the physical periapsis state and setting $P_0=2\bar z_0 p_0$. We
verified these maps and the canonical initial conditions reconstruct the physical
state to machine precision, and that on the pure Kepler problem at
$e=0.9$ a fixed-step velocity-Verlet integration gives a maximum relative energy
error of $1.46\times10^{-7}$ in Levi--Civita coordinates versus
$6.16\times10^{-3}$ in Cartesian --- a factor of $4.2\times10^{4}$ --- with
$u(s)$ a pure sinusoid, confirming the harmonic reduction.

\paragraph{Treatment of the orbit energy $E$.}
The regularized Hamiltonian~\eqref{eq:gamma} contains the orbit energy $E$ as
the oscillator frequency parameter. Throughout, we compute $E=H(q_0,p_0)$
\emph{analytically at the initial state}, including the analytic perturbation,
and hold it fixed on the energy shell. This is a fixed-shell reduction, not the
full extended-phase-space construction ($E=-p_t$), and it means $E$ carries
information a fully self-contained learned model would not have. We flag this as
a scope limitation rather than a solved design choice; the cleaner
extended-phase-space formulation is future work.

\paragraph{Gray-box decomposition.}
Following standard physics-informed practice, we do not learn gravity: the
harmonic Kepler core in Eq.~\eqref{eq:gamma} is hardcoded and exact, and a
network learns only the residual $U_{\mathrm{res}}$. Because the network outputs
a \emph{scalar} potential and the force is its gradient, the force is curl-free
by construction. Our \emph{analytic} experiments (Table~\ref{tab:arch},
Figure~\ref{fig:graybox}, exact-feature ladder) integrate the separable
Eq.~\eqref{eq:gamma} with a fixed-$\Delta s$ symplectic Störmer--Verlet splitting;
the \emph{learned} $2\times2$ ablation (Section~\ref{sec:ablation}) instead uses
explicit RK4 for all arms, which is non-symplectic --- we keep the two integrator
choices distinct and label each where it is used.

\paragraph{Branch/gauge and metrics.}
The planar map $q=z^2$ is $2{:}1$ ($(u,w)\sim(-u,-w)$). We select the initial
square-root branch by $z_0=\sqrt{q_0}$ (principal branch) and continue it
smoothly along each trajectory via $\texttt{np.unwrap}(\theta)/2$ on the polar
angle; the analytic residual~\eqref{eq:ures} is gauge-invariant by construction,
whereas a generic MLP is not (Section~\ref{sec:negative}). Unless stated
otherwise, ``$\max\lvert\Delta E/E\rvert$'' is the maximum over the rollout of
the relative physical-energy error $\lvert H(q,p)-H_0\rvert/\lvert H_0\rvert$. We
define survival operationally: a rollout \emph{survives} if it reaches its
prescribed numerical endpoint (the fixed step count over its prescribed horizon)
without producing a non-finite state (NaN/Inf) and without $\lVert q\rVert>60a$;
otherwise it is counted as \emph{diverged}. This endpoint-based
definition is unambiguous for the precessing perturbed trajectories, the $2S_0$
fictitious-time arms, and rollouts whose $s\mapsto t(s)$ map is distorted by
$\mathcal{O}(1)$ energy error; whether a physical periapsis return actually
occurred is reported separately where relevant. ``Fit residual'' in
Table~\ref{tab:ladder} is the maximum absolute error of the fitted potential over
the training samples.
Multi-seed learned results report the median over seeds of the per-run
$\max\lvert\Delta E/E\rvert$.

\paragraph{Protocol.}
We use two training grids. The learned-residual experiments
(Sections~\ref{sec:arch},~\ref{sec:negative}) train on
$a\in\{0.7,1.0,1.3\}\times e\in\{0.90,0.93,0.96\}$ and are evaluated on held-out
orbits up to $e=0.99$, so the highest-eccentricity test points are mild
\emph{extrapolation} beyond the training range. The in-distribution ablation
(Section~\ref{sec:ablation}), designed specifically to remove that
extrapolation confound, trains on the extended grid
$a\in\{0.7,1.0,1.3\}\times e\in\{0.90,0.93,0.96,0.99\}$ and tests on held-out
orbits \emph{interior} to $e\in[0.90,0.99]$, so its generalization is
interpolation. Networks are tanh MLPs (hidden width 32--48) trained with Adam;
the analytic and exact-feature experiments (Sections~\ref{sec:arch},
\ref{sec:negative}) use $N=400$ steps over the nominal LC horizon $S_0$, while the
learned $2\times2$ ablation (Table~\ref{tab:ablation2x2}) uses $N=200$, with its
$s$-arms spanning $2S_0$ and its uniform-$t$ arms spanning $T_0$; all learned
results are reported over $5$--$8$ random seeds. The reported metric is the maximum relative
energy error $\max_t\lvert H(t)-H_0\rvert/\lvert H_0\rvert$ along the rollout.

\section{The architecture result: a regularized discretization baseline}
\label{sec:arch}
We first remove the learning question and ask what the \emph{architecture} can
do, using the analytic residual $U_{\mathrm{res}}=r\,V_{\mathrm{pert}}$ so that
force error is zero. With a symplectic integrator on Eq.~\eqref{eq:gamma}:

\begin{table}[h]
\centering
\begin{tabular}{lccc}
\toprule
 & Cartesian & Levi--Civita & Levi--Civita\\
eccentricity $e$ & (horizon $T_0$) & (nominal $S_0$, $t{\approx}0.8T_0$) & (matched $t{=}T_0$)\\
\midrule
0.915 & $1.44$ & $\mathbf{2.0\times10^{-5}}$ & $\mathbf{3.1\times10^{-5}}$\\
0.945 & $1.31\times10^{1}$ & $\mathbf{2.1\times10^{-5}}$ & $\mathbf{3.2\times10^{-5}}$\\
0.975 & $2.2\times10^{2}$ & $\mathbf{2.1\times10^{-5}}$ & $\mathbf{3.3\times10^{-5}}$\\
0.990 & $6.4\times10^{3}$ & $\mathbf{2.1\times10^{-5}}$ & $\mathbf{3.3\times10^{-5}}$\\
\bottomrule
\end{tabular}
\caption{Maximum relative energy error, analytic residual, symplectic
Störmer--Verlet, $N=400$ steps (equal force-evaluation budget across all three
columns). \emph{Cartesian}: $N=400$ over the physical horizon $T_0=2\pi a^{3/2}$.
\emph{Levi--Civita (nominal)}: $N=400$ over the nominal fictitious horizon
$S_0=\pi/\sqrt{\lvert E\rvert/2}$, whose measured physical elapsed time is only
$t/T_0\approx0.79$--$0.81$ once the perturbation is present. \emph{Levi--Civita
(matched)}: $N=400$ over the fictitious endpoint $S_\ast$ chosen so
$t(S_\ast)=T_0$ ($S_\ast/S_0\approx1.24$, i.e.\ slightly larger $\Delta s$), the
decisive equal-horizon \emph{and} equal-budget control. The regularized
formulation holds ${\sim}(2$--$3)\times10^{-5}$ through $e=0.99$ under both
horizon conventions, while the Cartesian arm degrades by orders of magnitude at
the same budget and physical horizon. Symplecticity alone does not rescue
Cartesian coordinates --- fixed-step symplectic integrators still require
vanishingly small steps near periapsis. Physical time is accumulated by
trapezoidal quadrature of $dt=r\,ds$ along the rollout.}
\label{tab:arch}
\end{table}

\paragraph{The ${\sim}10^{-5}$ value is a discretization baseline, not a ceiling.}
The regularized Hamiltonian is integrated at a fixed step in \emph{fictitious}
time $s$ (with $dt=r\,ds$), so the relevant convergence is second-order in the
fictitious-time step, $\mathcal{O}(\Delta s^2)$, equivalently $\mathcal{O}(N^{-2})$
for $N$ steps per nominal Kepler horizon $S_0$ (see ``Fictitious-time horizon''). For pure Kepler, one physical orbit corresponds to
$s=\pi/\sqrt{\lvert E\rvert/2}$, \emph{half} the lifted oscillator's period,
because the Levi--Civita map $q=z^2$ identifies $z$ and $-z$ with the same
physical state (Section~\ref{sec:setup}). For pure Kepler, $S_0$ maps to one
physical orbit; in the perturbed experiments $N$ denotes steps over the explicitly
stated fictitious-time horizon ($S_0$ here), with the actual physical elapsed time
reported separately ($t/T_0\approx0.8$; Section~\ref{sec:setup}). A sweep at
$e=0.9$ confirms the order (velocity-Verlet):
$\max\lvert\Delta E/E\rvert =
3.22\times10^{-4},\,8.05\times10^{-5},\,2.01\times10^{-5},\,5.03\times10^{-6},\,
1.26\times10^{-6},\,3.14\times10^{-7}$ for $N=100,200,400,800,1600,3200$
(a factor ${\approx}4$ per doubling, log--log slope $-2.00$;
Figure~\ref{fig:ladder}, left). The ${\sim}2.1\times10^{-5}$ we quote at $N=400$
(Table~\ref{tab:arch}, measured at $e=0.975$--$0.99$; the sweep gives
$2.01\times10^{-5}$ at $e=0.9$) is therefore a \emph{resolution-dependent
discretization baseline} at that step count, not an
architectural limit: it falls ${\approx}4\times$ per step-count doubling with no
plateau. We use ``baseline'' rather than ``floor'' throughout to avoid implying
refinement stops helping --- it does not --- and ${\sim}10^{-5}$ at this step
count is not by itself high-precision celestial propagation.

\begin{figure}[h]
\centering
\includegraphics[width=0.95\textwidth]{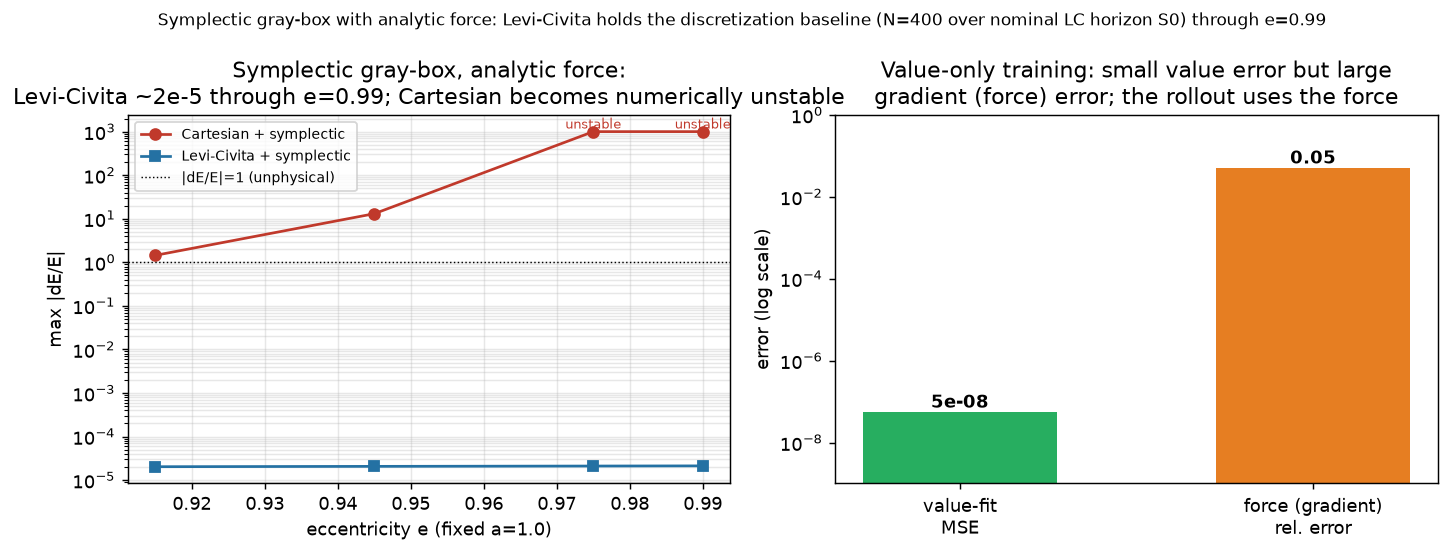}
\caption{Left: with the correct (analytic) force, the Levi--Civita gray-box is
flat at ${\sim}2\times10^{-5}$ to $e=0.99$ while the Cartesian rollout becomes
numerically unstable (finite but catastrophic energy error, growing to
$\mathcal{O}(10^3)$ at $e=0.99$; Table~\ref{tab:arch}). Right: why value-trained
residuals fail --- tiny value error, large force (gradient) error.}
\label{fig:graybox}
\end{figure}

\medskip\noindent
Figure~\ref{fig:graybox} also records a methodological point we had to debug: a
residual potential trained to near-perfect \emph{value} accuracy
(MSE $5\times10^{-8}$) can still have $5\%$ median (and $138\%$ maximum)
\emph{gradient} error, and the symplectic integrator faithfully integrates the
wrong force. Potential-value accuracy alone is thus insufficient to ensure
accurate forces or rollouts; derivative-aware (Sobolev) objectives are an
important control but were not, on their own, sufficient here.

\section{Rollout-survival advantage under matched sampling}
\label{sec:ablation}
A natural objection is that the advantage is really \emph{adaptive
time-stepping} (the Sundman clock concentrates steps at periapsis), which the
recent neural-symplectic literature already provides by other means. We ran a
$2\times2$ ablation over
$\{\text{Cartesian},\text{Levi--Civita}\}\times\{\text{uniform-}t,\text{Sundman-}ds\}$
at matched step budget $N=200$ (Table~\ref{tab:ablation2x2}). The cells differ in
their state equations and step variable but \emph{not} in the integrator: all
four use the same explicit RK4 scheme (Section~\ref{sec:setup}, ``Integration
protocol''), so none is symplectic: using the same RK4 method removes integrator
order and symplecticity as explanations, and the experiment compares the complete
coordinate-and-clock formulations, including their different analytic cores and
learned targets. The Cartesian arms integrate
$\dot q=p,\ \dot p=-\mu q/\lVert q\rVert^3-\nabla U_\theta$; the LC uniform-$t$ arm
re-expresses the lifted field in physical time via $d/dt=(1/r)\,d/ds$, which
reintroduces the $1/r$ scaling; and the LC Sundman-$ds$ arm advances the separable
field of Eq.~\eqref{eq:gamma} in fictitious time (still with RK4, not a symplectic
splitting). ``Matched'' means equal integrator, equal step count, and equal
integrator output points across arms. The Levi--Civita formulation outperforms
the Cartesian one even under a uniform physical-time clock (adaptivity removed),
although this comparison still conflates the coordinate change, the analytic
oscillator core, the altered learned target, and the oracle energy
(Section~\ref{sec:ablation}, ``$E$-information confound''); we do not attribute
the effect to the coordinate change in isolation. The Sundman clock in a
\emph{learned} rollout did not help and could hurt.

\begin{table}[h]
\centering
\begin{tabular}{lllcc}
\toprule
representation & clock & numerical scheme & finite runs & median max $\lvert\Delta E/E\rvert$\\
\midrule
Cartesian & uniform $t$ & RK4 (4th-order, non-sympl.) & $4/4$ & $\approx25$\\
Cartesian & Sundman $s$ & RK4 (4th-order, non-sympl.) & $0/4$ & diverged (all seeds)\\
Levi--Civita & uniform $t$ & RK4 (4th-order, non-sympl.) & $4/4$ & $\approx7.0$\\
Levi--Civita & Sundman $s$ & RK4 (4th-order, non-sympl.) & $4/4$ & $\approx7.1$\\
\bottomrule
\end{tabular}
\caption{Full $2\times2$ coordinate$\times$clock ablation, test orbit $a=1.0,e=0.9$
(periapsis start) --- at the low edge of the training grid
$e\in\{0.90,0.93,0.96\}$, i.e.\ \emph{in-distribution}, not OOD; $4$ seeds per arm.
All four arms use the \emph{same} explicit RK4 integrator (Section~\ref{sec:setup},
``Integration protocol''): a matched-order, matched-scheme comparison in which no
arm has an integrator-symplecticity advantage. The differences are thus not
integrator-order or symplecticity artifacts; they reflect the complete
coordinate-and-clock formulations, including their analytic cores, learned targets,
and oracle-$E$ asymmetry (see ``$E$-information confound''). The $s$-arms ran the
full lifted oscillator period ($2S_0$) at
$N=200$ versus the nominal period for the uniform-$t$ arms; because learned energy
errors distort the $s\mapsto t(s)$ map, we do not interpret this as a matched or
longer \emph{physical} horizon (Section~\ref{sec:setup}, ``Fictitious-time
horizon''), and read the table as a boundedness comparison under the stated
horizons. ``Finite runs'' counts seeds with a bounded rollout ($\lVert q\rVert<60a$); the
median is over finite runs. The finite-rollout advantage persists under both
tested clocks, while \emph{neither} learned LC arm achieves accurate energy
conservation (medians ${\sim}7$, i.e.\ $700\%$ energy error --- survival, not
accuracy). Central values are to two significant figures; the finite-run counts
are the robust, seed-independent part of this result.}
\label{tab:ablation2x2}
\end{table}

\paragraph{Integration protocol and discrete maps.}
Each arm advances a fixed step count over its prescribed horizon: $T_0=2\pi a^{3/2}$
for the uniform-time arms and $2S_0$ for the Sundman-time arms (Table~\ref{tab:ablation2x2}
uses $N=200$; the analytic experiments of Sections~\ref{sec:arch}, \ref{sec:negative}
use $N=400$ over $S_0$). \emph{All four learned ablation arms use the same explicit classical
fourth-order Runge--Kutta (RK4) integrator} --- $k_1=f(y)$, $k_2=f(y+\tfrac{h}{2}k_1)$,
$k_3=f(y+\tfrac{h}{2}k_2)$, $k_4=f(y+hk_3)$, $y\mathrel{+}{=}\tfrac{h}{6}(k_1+2k_2+2k_3+k_4)$
--- applied to each arm's first-order vector field $f$ with step $h$ ($\Delta t$
for the uniform-$t$ arms, $\Delta s$ for the Sundman-$s$ arms). The vector fields
differ by arm: Cartesian uniform-$t$ uses $\dot q=p,\ \dot p=-\mu q/r^3-\nabla U_\theta$;
LC Sundman-$s$ uses the separable $\Gamma$ field of Eq.~\eqref{eq:gamma}
($\dot z=P/4,\ \dot P=-2\lvert E\rvert z-\nabla_z U_{\mathrm{res}}$); Cartesian
Sundman-$s$ uses $q'=r\,p,\ p'=r(-\mu q/r^3-\nabla U_\theta)$; and LC uniform-$t$
uses $\dot z=P/4r,\ \dot P=-r^{-1}\big(2\lvert E\rvert z+\nabla_z U_{\mathrm{res}}\big)$,
i.e.\ the physical-time re-expression of the \emph{same} lifted field as the
Sundman-$s$ arm (equivalently $V_{\mathrm{LC}}\equiv\lvert E\rvert r+U_{\mathrm{res}}$,
which retains the harmonic Kepler core, not only the residual). RK4 is fourth-order
accurate but \emph{not} symplectic for any arm (including LC-Sundman, whose
separable Hamiltonian \emph{admits} a symplectic splitting we do not use here),
so this is a matched-order, matched-scheme, matched-step-count comparison in which
no arm has an integrator-symplecticity advantage --- the differences are not
integrator-order or symplecticity artifacts, but reflect the complete
coordinate-and-clock formulations (analytic cores, learned targets, and oracle-$E$
asymmetry included). RK4 uses four
force evaluations per step. (The \emph{analytic} experiments of
Table~\ref{tab:arch} and Figure~\ref{fig:graybox} instead use a symplectic
Störmer--Verlet splitting of the separable $\Gamma$, which is where the
$\mathcal{O}(\Delta s^2)$ discretization baseline is measured; the learned and
analytic tracks therefore use different integrators, stated here to avoid
conflation.)

\paragraph{Fictitious-time horizon.}
For the \emph{unperturbed} Kepler problem, one physical orbit corresponds to
$s=S_0\equiv\pi/\sqrt{\lvert E\rvert/2}$ --- \emph{half} the lifted oscillator
period $2\pi/\sqrt{\lvert E\rvert/2}$ --- because $q=z^2$ makes $z$ and $-z$ the
same physical state, so the physical trajectory completes a full orbit in half an
oscillator period. These are two different clocks: the fictitious-time horizon is
$S_0=2\pi\sqrt a$, and integrating $dt/ds=r$ over this interval yields the physical
Kepler period $T_0=2\pi a^{3/2}$ for pure Kepler motion (verified numerically: the
pure-Kepler physical state first returns at $s=S_0$, with $t(S_0)=T_0$). We use
$S_0$ as the
\emph{nominal (unperturbed) Kepler horizon} for the analytic experiments
(Table~\ref{tab:arch}, Figure~\ref{fig:graybox}, and the exact-feature ladder);
all step counts $N$ are steps over this nominal horizon.
\emph{The half-period identity is exact only for pure Kepler.} With the quadrupole
perturbation ($U_{\mathrm{res}}\neq0$) the lifted motion is no longer a pure
harmonic oscillator, so integrating to $S_0$ need not correspond to exactly one
completed physical orbit. We therefore report the actual physical elapsed time
$t_{\mathrm{end}}=\int_0^{S_0}r\,ds$ (accumulated by trapezoidal quadrature of
$dt=r\,ds$ along the rollout) relative to the nominal period $T_0=2\pi a^{3/2}$:
for the analytic LC gray-box at $a=1,N=400$ we measure
$t_{\mathrm{end}}/T_0=0.81,0.80,0.79,0.79$ at $e=0.915,0.945,0.975,0.99$ --- i.e.\
the $S_0$ horizon covers ${\sim}80\%$ of a nominal period once the perturbation is
present. To settle the resulting fairness question --- equal nominal $N$ gives
unequal physical horizon --- we run the decisive \emph{equal-horizon, equal-budget}
control: choose the fictitious endpoint $S_\ast$ with $t(S_\ast)=T_0$
($S_\ast/S_0\approx1.24$), divide it into exactly $N=400$ fixed-$\Delta s$ steps,
and compare against $400$ Cartesian steps over $T_0$. $S_\ast$ was determined
offline by a high-resolution reference solve followed by bisection on the total
fictitious interval until the trapezoidally accumulated $t(S_\ast)$ matched $T_0$;
the equal-budget statement refers to the final compared rollouts, each containing
$400$ force-gradient evaluations after initialization (the offline calibration cost
is excluded, and we match evaluation count, not FLOPs or wall time). At this
matched physical horizon and matched force-evaluation budget the LC baseline is
$3.1$--$3.3\times10^{-5}$ (Table~\ref{tab:arch}, third column) --- slightly larger
than the nominal $2.1\times10^{-5}$ because the $24\%$ longer fictitious interval
means slightly larger $\Delta s$, but still approximately $4.7$--$8.3$ orders of
magnitude below the Cartesian arm at the same budget. The LC advantage is therefore not an artifact
of physical-horizon or rollout force-evaluation accounting.

The learned $2\times2$ ablation (Table~\ref{tab:ablation2x2}) used the \emph{full}
lifted oscillator period ($2S_0$) for its $s$-arms at $N=200$, against the nominal
period for the uniform-$t$ arms. Because the learned rollouts carry
$\mathcal{O}(1)$ energy error, the mapping $s\mapsto t(s)=\int r_\theta\,ds$ is
itself distorted, so we do \emph{not} claim this corresponds to a matched or
longer physical duration; the $s$-arms simply use a longer nominal fictitious-time
horizon, and Table~\ref{tab:ablation2x2} should be read as a boundedness
comparison under the stated horizons, not a controlled equal-physical-time test.
Energy error is evaluated at every integrator output point in the native clock (we
report the per-run maximum, not an interpolation to matched physical times);
because the two clocks place output points differently, the comparison is of
boundedness and worst-case energy drift, not of pointwise trajectory agreement ---
the latter requires the trajectory-level metrics we defer to future work.

\begin{figure}[h]
\centering
\includegraphics[width=0.95\textwidth]{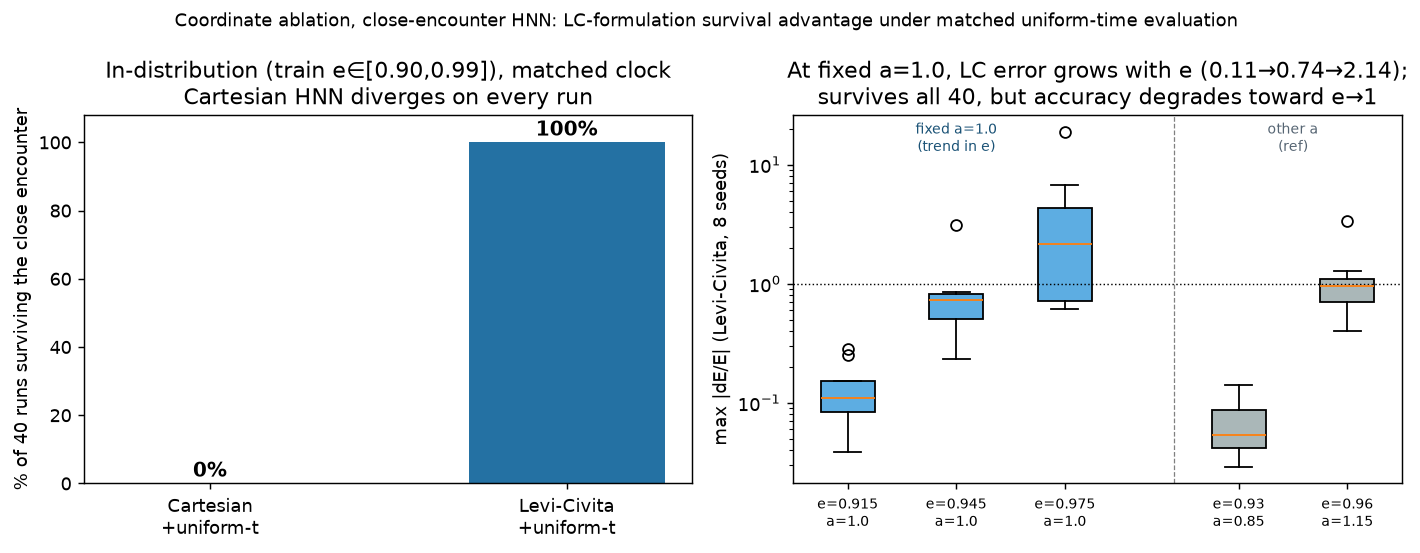}
\caption{In-distribution ablation, matched clock. Left: Cartesian $0/40$
survival vs Levi--Civita $40/40$. Right: at fixed $a=1.0$ the surviving
Levi--Civita error grows with $e$ (grey boxes are the two off-$a$ reference
orbits, not part of the trend).}
\label{fig:ablation}
\end{figure}

We then removed the out-of-distribution confound by training across
$e\in[0.90,0.99]$ and testing on held-out interior orbits, over 8 seeds
$\times$ 5 orbits (Figure~\ref{fig:ablation}):
the Cartesian learned integrator diverged on \textbf{all 40} runs; the
Levi--Civita one on \textbf{none}. This is the strongest, most qualitative form
of the result: the tested Cartesian models failed to approximate the
near-periapsis field accurately enough to integrate through a high-eccentricity
close encounter at the chosen capacity, data distribution and training budget,
whereas the regularized network survives every time. The two arms are matched on
physical-time sampling (of training observations, integrator output points, and
force evaluations alike), network evaluations, parameter count, normalization and
training examples, and both use the same explicit RK4 integrator. This is a
\emph{matched uniform-time} comparison: the LC arm evolves the physical-time
re-expression of the lifted regularized vector field
(the first-order ODE system $d/dt=(1/r)\,d/ds$, integrated with fourth-order RK4),
and the Cartesian arm its physical-coordinate counterpart;
neither is the canonical fixed-$\Delta s$ symplectic rollout of
Eq.~\eqref{eq:gamma}. The advantage is therefore of the LC \emph{formulation}
under a matched clock, not of the coordinate change in isolation (it still
conflates the oracle $E$, the analytic core, the transformed target, and the state
representation; Section~\ref{sec:ablation}, ``$E$-information confound'').

\paragraph{Survival is much stronger than accuracy.}
We stress that this is a statement about \emph{rollout survival}, not trajectory
accuracy: many surviving LC rollouts carry relative energy errors of order $1$--$10$
(Table~\ref{tab:ablation2x2}), i.e.\ finite but physically poor. A bounded
trajectory with $700\%$ energy error demonstrates numerical survivability, not
successful learned dynamics. We therefore do not claim the regularized learned
model \emph{simulates} close encounters accurately; the defensible headline is
that regularization converts guaranteed Cartesian divergence into bounded (if
inaccurate) motion. Trajectory-level metrics --- position error at matched
physical times, periapsis distance and timing error, orbital-phase and
angular-momentum drift --- are the right way to quantify accuracy and are left to
future work; without them, ``survival'' should not be read as ``accurate
simulation.''

\paragraph{An $E$-information confound.}
We deliberately state the result as a survival \emph{advantage under matched
sampling and compute}, not as a clean isolation of the coordinate change. The
regularized Hamiltonian~\eqref{eq:gamma} uses the exact orbit energy $E$ as its
oscillator-frequency parameter (Section~\ref{sec:setup}), and the Cartesian arm
receives no equivalent orbit-specific scalar. Across trajectories with varying
$a$, that exact $E$ carries substantial global information about each orbit, so
the $40/40$ vs $0/40$ outcome conflates the coordinate transform, the exact
Kepler-core structure, an orbit-specific oracle energy, the altered target and
the altered numerical flow. Disentangling these requires giving the Cartesian
arm the same $E$ (or a learned/noisy estimate), or the extended-phase-space
construction $E=-p_t$; we have not run that control and flag it as the main
threat to the coordinate reading of this section. Survival, moreover, is not
accuracy --- with a non-symplectic rollout the regularized model's error still
grew with $e$ (right panel), foreshadowing Section~\ref{sec:negative}.

\section{The exact-feature optimization gap is a basis-conditioning trade-off}
\label{sec:negative}
\begin{figure}[h]
\centering
\includegraphics[width=0.8\textwidth]{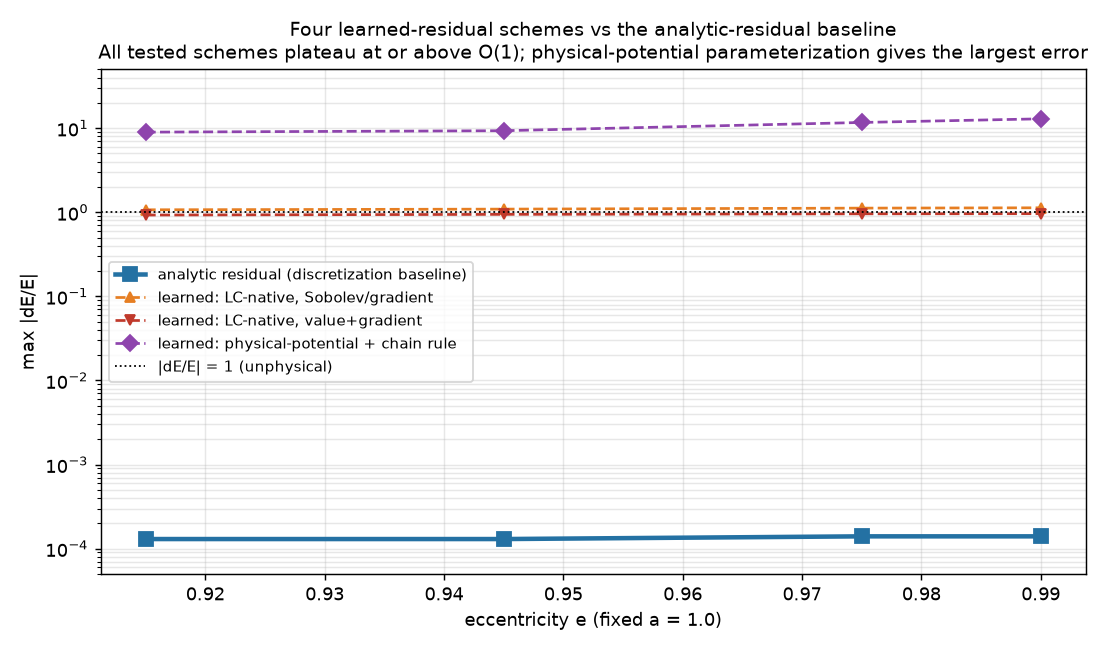}
\caption{Four learned-residual schemes against the analytic-residual baseline. All
plateau or diverge at $\mathcal{O}(1)$; the physical-potential chain-rule variant
is the worst.}
\label{fig:routes}
\end{figure}

The architecture reaches the discretization baseline with the analytic residual. Does a
\emph{learned} residual? Across four neural schemes, no (Table~\ref{tab:learned},
Figure~\ref{fig:routes}):

\begin{table}[h]
\centering
\begin{tabular}{lc}
\toprule
approach (Levi--Civita, symplectic) & max $\lvert\Delta E/E\rvert$ at $e=0.975$\\
\midrule
analytic residual (discretization baseline, $N{=}400$) & $\mathbf{2.1\times10^{-5}}$\\
learned, potential-value regression & diverges\\
learned, gradient/Sobolev matching & $1.12$\\
learned, balanced value$+$gradient & $0.96$\\
learned, physical potential $+$ chain-rule force & $11.7$\\
\bottomrule
\end{tabular}
\caption{Every learned scheme plateaus at or above $\mathcal{O}(1)$ energy error,
roughly five orders of magnitude short of the analytic-residual baseline
($2.1\times10^{-5}$). Gradient
(Sobolev) training reduced force error from $5\%$ median / $138\%$ max to
$0.35\%$ / $7.4\%$, but this did not translate into low energy error.}
\label{tab:learned}
\end{table}

\paragraph{Mechanism, partially characterized.}
Symplectic integrators do not exactly conserve the integrated Hamiltonian; by
backward-error analysis they nearly conserve a modified (shadow)
Hamiltonian~\cite{hairer2006}. The true energy drift decomposes approximately as
$\Delta H_{\mathrm{true}}\approx\Delta(H_{\mathrm{true}}-H_\theta)+\Delta
H_\theta^{\mathrm{num}}$, a model-error term plus a numerical (shadow) term.
Empirically the along-trajectory variation of the residual-potential error
$U_\theta-U_{\mathrm{res}}$ correlates with the observed energy error at
$\rho=0.94$ (peak-to-peak value error ${\approx}2.4$ vs energy error
${\approx}1.2$), which is evidence --- not proof of exclusive causation --- that
inaccurate potential \emph{differences} between dynamically visited states
dominate the gap. The learned force is the gradient of a scalar potential, hence
curl-free by construction; the issue is not path-dependence but inaccurate
potential differences. Among training variants, Sobolev/gradient matching reduced
force error (median $5\%\to0.35\%$) without fixing energy; a physical-coordinate
parameterization with chain-rule force reconstruction (verified to $10^{-17}$)
did \emph{worse}; a uniform multiplicative force bias in the analytic rollout
produced only small energy error ($0.3\%\to3\times10^{-4}$); and value
re-weighting barely moved the plateau ($1.07\to0.93$).

\paragraph{Exact-feature control: representation vs.\ optimization.}
An exact-feature control separates representational capacity from optimization
failure. In Levi--Civita coordinates the residual is
\begin{equation}
U_{\mathrm{res}}=r\,V_{\mathrm{pert}}
=A\big(u^6-3u^4w^2-3u^2w^4+w^6\big),
\label{eq:ures}
\end{equation}
a four-monomial degree-6 polynomial (degree-2 in physical $(x,y)$). A
\emph{linear least-squares} fit on exact degree-6 monomial features reproduces
$U_{\mathrm{res}}$ to $9\times10^{-16}$ and its symplectic rollout reaches the
discretization baseline ($2.1\times10^{-5}$ at $e=0.975$); a degree-2 fit in physical
coordinates does the same. This control supports the internal consistency of the
shell-initialization, reconstruction, force-conversion, and energy-evaluation
components exercised by the experiment, while showing that the target is exactly
representable. So the exact-feature gap is about optimization rather than
representational capacity, and
the symplectic rollout amplifies small potential-difference errors into
$\mathcal{O}(1)$ energy drift.

\begin{figure}[h]
\centering
\includegraphics[width=0.95\textwidth]{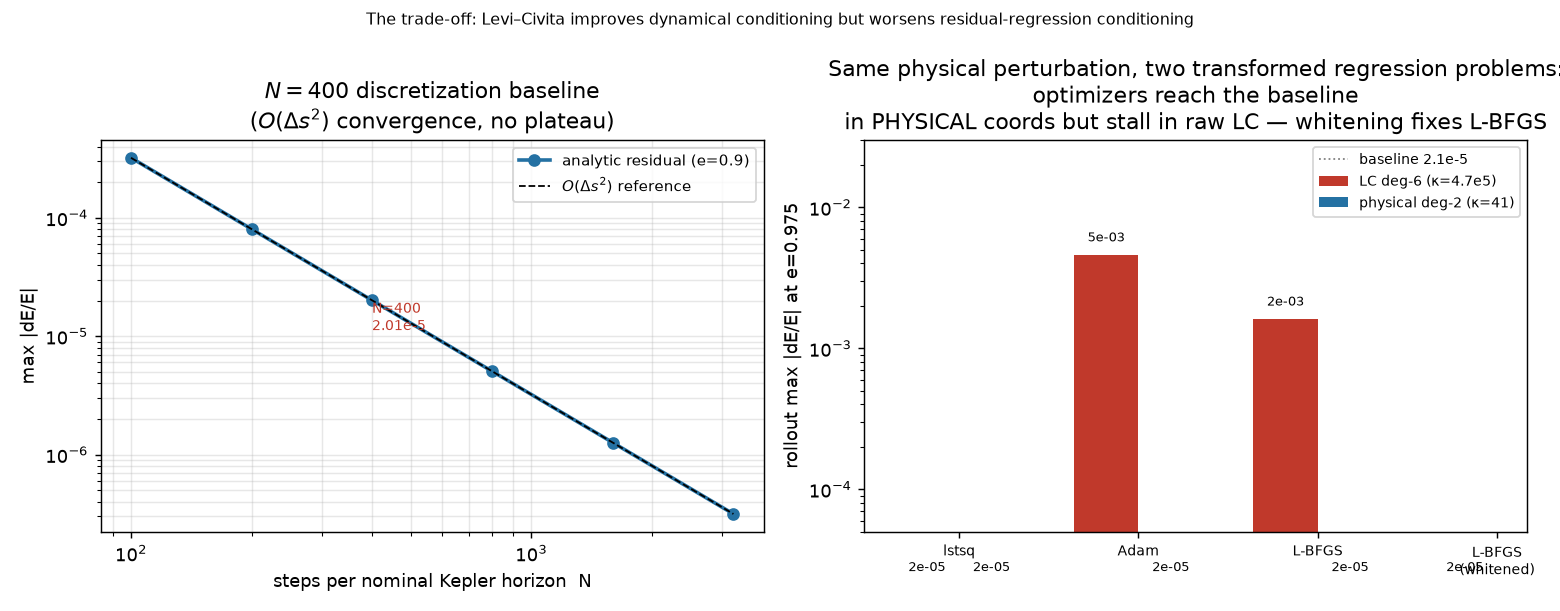}
\caption{Left: the ${\sim}2\times10^{-5}$ value is a discretization baseline converging as
$\mathcal{O}(\Delta s^2)$ (fictitious-time step), not a ceiling. Right: the
conditioning trade-off --- for the same physical perturbation in the two
coordinate systems, iterative optimizers
reach the baseline in the well-conditioned physical basis ($\kappa=41$) but stall
in the ill-conditioned Levi--Civita basis ($\kappa=4.7\times10^{5}$); whitening
the LC basis restores baseline-level fitting for L-BFGS (two iterations).}
\label{fig:ladder}
\end{figure}

\paragraph{The optimization difficulty is coordinate-and-basis-induced conditioning.}
We separate two notions of conditioning:
\emph{dynamical/numerical} conditioning (smoothness and integrability of the
near-encounter trajectory), which regularization \emph{improves}, and
\emph{regression/optimization} conditioning (the geometry of fitting the residual
in a given basis), which we now show it \emph{worsens}. For the same underlying
physical perturbation in the two coordinate systems,
the tested optimizers behave very differently
(Table~\ref{tab:ladder}, Figure~\ref{fig:ladder} right):
\begin{itemize}
\item \textbf{Physical} degree-2 basis ($\kappa(X)=41$): Adam and L-BFGS both
reach the baseline ($2.1\times10^{-5}$), L-BFGS in $14$ iterations. Iterative
optimization is trivial here.
\item \textbf{Levi--Civita} degree-6 basis ($\kappa(X)=4.7\times10^{5}$,
$\kappa(X^\top X)\approx2.2\times10^{11}$): L-BFGS runs $7813$ iterations and
still stalls at $1.6\times10^{-3}$; raw-basis Adam (lr~$10^{-2}$, $2\times10^4$ it.)
plateaus at $4.6\times10^{-3}$; the tanh-MLP is worst at ${\sim}1.1$. Column
normalization (which only rescales columns, leaving the four monomials
$u^6,u^4w^2,u^2w^4,w^6$ nearly collinear) lowers $\kappa$ to $1.9\times10^{4}$ but
does not help (normalized-basis Adam plateaus at $2.1\times10^{-2}$) --- consistent
with a highly correlated, not merely badly scaled, basis.
\item \textbf{Whitened} Levi--Civita ($X=QR$, optimize in the orthonormal $Q$
basis): \emph{L-BFGS} reaches the baseline in \textbf{$2$ iterations}, so
orthogonalization removes the raw-basis obstacle for the quasi-Newton optimizer.
\emph{Adam} reaches \emph{none} of the baselines at any tested configuration
(rollout $3.0\times10^{-3}$ at $\mathrm{lr}=3\times10^{-3}$, $6$k steps;
$1.0\times10^{-2}$ at $\mathrm{lr}=10^{-2}$, $2\times10^4$ steps), and both remain
${\gtrsim}100\times$ above the $2.1\times10^{-5}$ baseline. Because
matched-step gradient descent solves the whitened quadratic immediately (below),
the remaining Adam behavior is optimizer-specific rather than a conditioning
limitation; we do not diagnose it further. Write the whitened least-squares
objective explicitly as
\begin{equation}
L(\beta)=\tfrac{1}{2n}\lVert Q\beta-y\rVert_2^2,\qquad
\nabla^2 L=\tfrac{1}{n}Q^\top Q=\tfrac{1}{n}I,
\end{equation}
using $Q^\top Q=I$ ($\lVert Q^\top Q-I\rVert_F=9\times10^{-15}$, verified) --- a
perfectly isotropic bowl with all Hessian eigenvalues equal to $1/n$ and no
ill-conditioned directions. The one-step convergence then becomes transparent
rather than surprising: full-batch gradient descent from $\beta=0$ with step
$\eta=n$ gives $\beta_1=\beta^\star$ exactly, and we verify a relative coefficient
error $\lVert\beta_1-\beta^\star\rVert_2/\lVert\beta^\star\rVert_2=9\times10^{-17}$
in one step. Since a single matched-step gradient-descent step already solves the
whitened problem exactly, the objective is provably well-conditioned; Adam's
failure on it is therefore optimizer-specific and not a basis-conditioning effect.
Tuned Adam does \emph{not} reach the baseline at any tested rate
(${\gtrsim}100\times$ above it); only L-BFGS and matched-step GD reach it exactly.
We do not further diagnose the exact reason Adam fails on the well-conditioned
objective.
\end{itemize}
The exact-feature story thus narrows precisely: raw-basis conditioning explains
the raw-vs-whitened \emph{L-BFGS} behavior; it does \emph{not} explain Adam
(which fails even on the provably well-conditioned whitened objective, for reasons
we do not diagnose); and neither explains the MLP.
Why does the transform raise conditioning? The mechanism is a degree count: a
degree-$d$ physical polynomial generically becomes a polynomial of degree at most
$2d$ under $q=z^2$, and after multiplication by the Sundman factor
$r=\lvert z\rvert^2$ at most $2d+2$ (cancellations can lower it). For the present
quadratic perturbation ($d=2$) the resulting degree is exactly $6$
(Eq.~\eqref{eq:ures}), and the high-degree monomials $u^6,u^4w^2,u^2w^4,w^6$ are
strongly correlated over the sampled trajectories --- a mechanistic, not merely
empirical, account of the trade-off.

The conclusion is a \emph{trade-off}, not an artifact and not a barrier: the
nonlinear Levi--Civita transform that removes the collision singularity and
stabilizes the high-eccentricity rollout simultaneously raises the residual's
polynomial degree, so that under the sampled trajectories and the raw monomial
basis the regression becomes severely ill-conditioned and the tested
\emph{raw-basis} iterative configurations do not recover the least-squares
solution; orthogonalization combined with L-BFGS or matched-step gradient descent
does (exactly, to the baseline), while tuned Adam does not reach it. We retract the
earlier ``fundamental bottleneck'' claim: for the \emph{exact-feature} problem
the residual-learning difficulty is coordinate-and-basis-induced regression
conditioning, addressable by preconditioning/whitening, not intrinsic
nonrepresentability of the coordinates.

\paragraph{The small-MLP gap is separate and remains open.}
The exact-feature analysis explains the \emph{polynomial} optimizer gap; it does
\emph{not} establish that the $\mathcal{O}(1)$ tanh-MLP rollout error
(Table~\ref{tab:learned}) has the same cause. The MLP and the exact-feature model
differ in more than conditioning: the MLP has redundant hidden-unit symmetries
and a nonconvex loss, can fit values well while producing poor derivatives, and
lacks two inductive biases the exact basis has for free --- the correct
polynomial degree and the planar-LC gauge symmetry $U(-u,-w)=U(u,w)$ (the map
$q=z^2$ is $2{:}1$). We tested whether supplying the gauge symmetry helps, using
a symmetrized head $U_\theta^{\mathrm{inv}}(z)=\tfrac12[f_\theta(z)+f_\theta(-z)]$:
across seeds it did \emph{not} close the gap (median rollout $\mathcal{O}(1)$,
comparable to the plain MLP), so the MLP failure is not explained by the missing
gauge invariance alone. Whether preconditioning, a richer architecture, or
derivative-space training closes the MLP gap is unresolved and is the main open
question left by this study.

\begin{table}[h]
\centering
\small
\begin{tabular}{llcc}
\toprule
basis ($\kappa(X)$) & solver & fit resid.\ & rollout $\lvert\Delta E/E\rvert$, $e{=}0.975$\\
\midrule
physical deg-2 ($41$) & lstsq & $7\times10^{-17}$ & $\mathbf{2.1\times10^{-5}}$\\
physical deg-2 ($41$) & Adam & $6\times10^{-10}$ & $\mathbf{2.1\times10^{-5}}$\\
physical deg-2 ($41$) & L-BFGS ($14$ it.) & $9\times10^{-11}$ & $\mathbf{2.1\times10^{-5}}$\\
\midrule
LC deg-6 ($4.7\times10^{5}$) & lstsq & $9\times10^{-16}$ & $\mathbf{2.1\times10^{-5}}$\\
LC deg-6 ($4.7\times10^{5}$) & L-BFGS ($7813$ it.) & $1.7\times10^{-5}$ & $1.6\times10^{-3}$\\
LC deg-6 ($4.7\times10^{5}$) & Adam (lr $10^{-2}$, $2{\times}10^4$ it.) & $3.8\times10^{-5}$ & $4.6\times10^{-3}$\\
LC deg-6 (col.-normalized, $1.9\times10^{4}$) & Adam (lr $10^{-2}$, $2{\times}10^4$ it.) & $3.5\times10^{-4}$ & $2.1\times10^{-2}$\\
\midrule
LC deg-6 whitened ($X{=}QR$) & L-BFGS ($2$ it.) & $5.7\times10^{-15}$ & $\mathbf{2.1\times10^{-5}}$\\
LC deg-6 whitened ($X{=}QR$) & GD, $\eta{=}n$ ($1$ step) & $6.0\times10^{-15}$ & $\mathbf{2.1\times10^{-5}}$\\
LC deg-6 whitened ($X{=}QR$) & Adam (lr $3{\times}10^{-3}$, $6$k it.) & $1.8\times10^{-5}$ & $3.0\times10^{-3}$\\
LC deg-6 whitened ($X{=}QR$) & Adam (lr $10^{-2}$, $2{\times}10^4$ it.) & $5.9\times10^{-5}$ & $1.0\times10^{-2}$\\
LC (tanh-MLP, Table~\ref{tab:learned}) & Adam & --- & ${\sim}1.1$\\
\bottomrule
\end{tabular}
\caption{Exact-feature control ladder (rollout at $e{=}0.975$, nominal Kepler horizon $S_0$,
$N=400$). The residual is exactly representable in both coordinate systems, so a
direct least-squares solve reaches the discretization baseline $2.1\times10^{-5}$
everywhere. Iterative optimizers reach the baseline in \emph{physical} coordinates
($\kappa=41$) but stall in the raw \emph{Levi--Civita} basis
($\kappa=4.7\times10^{5}$, $\kappa(X^\top X)\approx2.2\times10^{11}$); column
normalization does not help, but orthogonalizing (whitening) the LC basis makes
the objective isotropic ($\lVert Q^\top Q-I\rVert=9\times10^{-15}$). On the
whitened basis, L-BFGS reaches the baseline in two iterations and full-batch
gradient descent at the matched step $\eta=n$ reaches it in a \emph{single} step
(relative coefficient error $9\times10^{-17}$); Adam does \emph{not} reach the
baseline at any tested rate (rollout $3.0\times10^{-3}$ at lr~$3\times10^{-3}$ and
$1.0\times10^{-2}$ at lr~$10^{-2}$, both ${\gtrsim}100\times$ above baseline),
confirming that the residual Adam gap is optimizer-specific on an already
well-conditioned objective (matched-step GD solves it in one step), not basis
conditioning (which GD and L-BFGS show is fully removed by whitening); we do not
diagnose the Adam failure further. The exact-feature difficulty is thus
coordinate-and-basis-induced regression conditioning, not nonrepresentability; the
small-MLP failure (last row) is separate and unresolved.}
\label{tab:ladder}
\end{table}

\paragraph{Whitening is train-only (no test leakage).}
The whitening transform is built from \emph{training} samples alone: we factor
the training design matrix $X=QR$ and define the transformed analytic feature map
$\widetilde\phi(z)^\top=\phi(z)^\top R^{-1}$, so that $\widetilde X=XR^{-1}=Q$.
Because $R^{-1}$ is applied to the \emph{analytic} feature vector, the whitened
potential and its derivatives are evaluated and differentiated at arbitrary
rollout states off the training grid; no test-trajectory samples enter the
construction of $R$.

\section{Related work (positioning)}
The components here are individually classical or published. Regularization of
the Kepler collision singularity (Levi--Civita~\cite{levicivita1920};
Kustaanheimo--Stiefel~\cite{stiefel1971}) and its use in $N$-body codes (chain
regularization, algorithmic/AR regularization) are foundational celestial
mechanics; the chain~\cite{mikkola1993} and algorithmic/AR~\cite{mikkola2008}
regularizations extend it to $N$-body hierarchies. Hamiltonian and symplectic
networks (HNN~\cite{greydanus2019}; Hamiltonian graph networks~\cite{sanchez2019};
SympNets~\cite{jin2020}), equivariant message passing (EGNN~\cite{satorras2021}),
gray-box/residual physics-informed learning, and learned coordinate transforms
for HNNs are all published; the last is closest to us:
\cite{unifying2022} learns an invertible map into coordinates where
Hamiltonian-style dynamics constraints are imposed, and \cite{koopman2024}
applies deep Koopman learning to perturbed two-body and restricted three-body
motion. We differ in using the \emph{analytic} Levi--Civita map (no learned
transform) with a learned scalar residual and a close-encounter symplectic
rollout. Two recent works from 2025--2026 are the closest adjacent prior art for our
individual components: NNPT~\cite{nnpt2025} learns residual corrections after
analytically subtracting an exact solution in a gravitational three-body setting
(the analytic-core-plus-learned-residual idea, but without Levi--Civita
coordinates, a scalar Hamiltonian residual, or regularized close-encounter
integration), and ATLAS-NN~\cite{atlasnn2026} learns a nonlinear temporal
reparameterization for HNNs (learned time warping, in contrast to our analytic
Sundman regularization). Residual learning around an analytic Kepler solution and
learned temporal reparameterization have thus both appeared independently; we study
their celestial-regularization analogue through an analytic LC lift, a scalar
regularized Hamiltonian residual, and close-encounter rollout. The recent
neural-symplectic line --- time-adaptive
HénonNets~\cite{henonnets2025} and related architectures --- is restricted to
separable Hamiltonians and adds \emph{learned} time-adaptivity, whereas
nonseparable symplectic networks~\cite{nssnn2020} handle nonseparable $H$ through
an augmented phase space; regularization instead supplies both a separable
(harmonic) core and canonical adaptivity classically. To our knowledge the
specific combination studied here --- an \emph{analytic} Levi--Civita map, a
learned smooth, nonsingular scalar residual on the regularized energy shell (bounded over the sampled compact domain), and a
close-encounter symplectic rollout --- has not been reported; we claim novelty for
that combination, not for any component in isolation.

\section{Limitations}
(0) \textbf{Scope of ``close encounter.''} We study high-eccentricity periapsis
passage in a two-body-dominated central potential --- \emph{near-collision
perturbed Kepler dynamics}. We do \emph{not} treat competing pairs, binary
exchange, close passage between independently moving massive bodies, or chart
selection; the title's ``close-encounter'' should be read in this restricted
sense.
(1) Two-body-dominated, single fixed regularization chart; the multi-body case
requires chart selection (chain regularization) that does not compose trivially
with a fixed graph and is untested here. (2) A single smooth analytic
perturbation; singular perturbations (e.g.\ $\propto1/r^2$) are only partially
softened by regularization and behaved worse in our tests. (3) No benchmark
against production integrators; the claim is representational, not a forward-problem
speed/accuracy claim. (4) We report a single conserved quantity (energy);
trajectory-level metrics (position/momentum error, periapsis distance and time,
orbital-phase and angular-momentum error) would strengthen the survival claim
and are left to future work. (5) The orbit energy $E$ is supplied analytically
on a fixed shell (Section~\ref{sec:setup}), so the learned experiments are not
fully self-contained; the extended-phase-space formulation ($E=-p_t$) is the
right fix. (6) The optimization/conditioning diagnosis is established for
polynomial bases; whether preconditioning or quasi-Newton optimization closes
the MLP gap, and whether the conditioning story survives richer perturbations,
are the immediate follow-ups.

\section{Conclusion}
In this planar perturbed-Kepler benchmark, the Levi--Civita formulation greatly
improves high-eccentricity rollout survival at a fixed evaluation budget
($40/40$ vs $0/40$) and allows an analytic residual to attain the expected
second-order discretization baseline in the fictitious-time step
($\mathcal{O}(\Delta s^2)$, ${\sim}2\times10^{-5}$ at $N=400$) through $e=0.99$, where
the Cartesian formulation becomes numerically unstable. The
transformed residual is exactly sparse and polynomial
(Eq.~\eqref{eq:ures}), ruling out intrinsic nonrepresentability. However, the
same transformation produces a highly correlated degree-6 regression problem on
the sampled trajectories ($\kappa(X)\approx4.7\times10^{5}$), and the tested
raw-basis LC optimization configurations fail to recover the direct least-squares
solution ---
whereas they recover it easily in physical coordinates ($\kappa=41$), and
whitening the Levi--Civita basis restores baseline-level fitting in two iterations.
These findings expose a \emph{trade-off} between two distinct notions of
conditioning --- dynamical/numerical conditioning of the rollout, which
regularization improves, and regression/optimization conditioning of the residual
fit, which the transformed raw feature parameterization worsens in this benchmark
--- rather than a fundamental barrier or a complete solution to learned
close-encounter dynamics. The practical implication is
scoped to what we tested: for residuals represented in an explicit transformed
polynomial basis, orthogonalization with a direct or quasi-Newton solver removes
the severe basis-conditioning effect (it fixed L-BFGS; it did not make Adam reach
the baseline under the tested configurations, and that Adam behavior is a separate
optimizer-level issue on an already well-conditioned objective). Whether analogous
preconditioning improves \emph{neural} residual models --- which remained at
$\mathcal{O}(1)$ here --- is open.

\paragraph{Reproducibility.}
All results are from self-contained numerical experiments (NumPy/JAX, double
precision). The perturbed Kepler dynamics, Levi--Civita transform, canonical
initial conditions, symplectic integrators, treatment of $E$, and all training
and control schemes are as specified in
Sections~\ref{sec:setup}--\ref{sec:negative}; per-experiment training grids are
stated in the Protocol paragraph of Section~\ref{sec:setup}. Key settings used
here: MLP residuals are two hidden layers of width $32$--$64$ with $\tanh$
activation; training uses full-batch Adam (learning rate $3\times10^{-3}$,
$\beta_1{=}0.9,\beta_2{=}0.999,\epsilon{=}10^{-8}$) for $6000$ steps; training
data are $300$ points sampled uniformly in fictitious time over the nominal LC
horizon $S_0$ for each training orbit on the stated $(a,e)$ grid ($3600$ samples);
multi-seed results use seeds $0$--$7$. The analytic and exact-feature experiments
use $N=400$ steps over the nominal one-orbit LC horizon $S_0=\pi/\sqrt{\lvert E\rvert/2}$;
the Table~\ref{tab:ablation2x2} $s$-arms use $N=200$ over the full lifted
oscillator period $2S_0$ (nominally two pure-Kepler periods), and the uniform-$t$
arms $N=200$ over the nominal Kepler period $T_0=2\pi a^{3/2}$; measured physical
elapsed times $t_{\mathrm{end}}/T_0$ are reported in Section~\ref{sec:ablation},
under ``Fictitious-time horizon.'' The matched-horizon control in
Table~\ref{tab:arch} uses $N=400$ over the calibrated endpoint $S_\ast$ with
$t(S_\ast)=T_0$.
Polynomial-basis controls use NumPy \texttt{lstsq}, SciPy L-BFGS-B, and
the full-batch GD/Adam runs reported in Section~\ref{sec:negative}. Because the
whitened-Adam behavior is hyperparameter-sensitive, the exact optimizer
schedules, stopping tolerances, batch handling, seed-level tables, and a pinned
code repository will accompany submission --- until then, the whitened-basis
result should be read as an optimizer diagnostic rather than a tuned benchmark.

\end{document}